\documentclass[aps,preprint]{revtex4}%
\usepackage{amsfonts}
\usepackage{amsmath}
\usepackage{amssymb}
\usepackage{graphicx}%
\begin{document}

\begin{center}
\textbf{Key principle of the efficient running, swimming, and flying }

.

by Valery B. Kokshenev

Submitted to EPL\ 11 June 2009, resubmitted 31 August 2009\ \ \ \ \ \ \ \ \ \ \ \ \ \ 

.
\end{center}

\textbf{Abstract. }Empirical observations indicate striking similarities among
locomotion in terrestrial animals, birds, and fish, but unifying physical
grounds are lacking. When applied to efficient locomotion, the analytical
mechanics principle of minimum action yields two patterns of mechanical
similarity via two explicit spatiotemporal coherent states. In steady
locomotory modes, the slow muscles determining maximal optimum speeds maintain
universal intrinsic muscular pressure. Otherwise, maximal speeds are due to
constant mass-dependent stiffness of fast muscles generating a uniform force
field, exceeding gravitation. Being coherent in displacements, velocities and
forces, the body appendages of animals are tuned to natural propagation
frequency through the state-dependent elastic muscle moduli.


\section{Introduction}

Although evolutionary biologists and comparative zoologists make wonderful
generalizations about the movements of terrestrial animals, birds, and fish of
different size [1-12], the fundamental physical principles underlying striking
similarities in distinct types of movement for organisms remain a challenge
[13]. Within the scope of the simplest pendulum model (stiff-legged
approximation), it has been demonstrated [14] that humans and other animals,
in contrast to human-made engines, accomplish efficient propulsion (maximum
power output at minimum power consumption) by tuning musculoskeletal system to
the resonant propagation frequency. Storing mechanical energy in elastic
oscillations of body parts and in pendulum oscillations of legs or other
appendages, animals thereby reduce the energy consumption [1,3], which is
minimal at the resonance conditions [14]. In this study, instead of searching
for uncovered principles of body mass effects in biology [5], or doing
in-depth analysis of equations of motion in pendulum [14], spring [7,8], or
vortex [15] approximations and other engineer constructive approaches [9], I
address the key principle of mechanics.

In analytical mechanics, the requirement of\emph{\ }minimum action between two
fixed points of the conceivable trajectory of an arbitrary isolated mechanical
system determines Lagrangian $\mathcal{L}(q,v)$, the function of
time-dependent coordinates $q(t)$ and instant velocities $v(t)=dq/dt$.\ The
most general property of a freely\emph{ }moving system is spatiotemporal
homogeneity implying that the multiplication of $\mathcal{L}$ on an arbitrary
constant does not affect the equations of motion, arising from $\mathcal{L}$.
This property, designated as a \emph{mechanical similarity} [16], permits one
to establish the major mechanical constraints without consideration of
equations of motion. Indeed, following Landau and Lifshitz [16], let
us\textrm{ }consider the uniform transformation of mechanical trajectories due
to linear changing of all coordinates $q\rightarrow aq$ and times
$t\rightarrow bt$, and hence velocities $v\rightarrow(a/b)v$, via arbitrary
coefficients $a$ and $b$. Let the potential energy change consequently through
a certain exponent $s$, \textit{i.e.}, $\mathcal{U}(aq)=a^{s}\mathcal{U}(q)$.
Being a quadratic function of velocities, the kinetic energy scales as
$\mathcal{K}(av/b)=\left(  a/b\right)  ^{2}\mathcal{K}(v)$. The requirement of
homogeneity of $\mathcal{L}(q,v)=\mathcal{K}(v)-\mathcal{U}(q)$ is
self-consistent when both the energies change similar, \textit{i.e.},
$(a/b)^{2}=a^{s}$ or $b=a^{1-s/2}$ . Thereby, the frictionless propagation of
a classical system obeys the scaling relationships imposed on all principal
mechanical characteristics: \emph{period} $T$, overall-system \emph{speed}
$V$, and \emph{force} \emph{amplitude} $F$, namely [16]%
\begin{equation}
T\backsim t\varpropto L^{1-s/2}\text{, }V\backsim v\varpropto L^{s/2}\text{,
and }F\varpropto L^{s-1}\text{.} \label{1A}%
\end{equation}
The seminal case $s=-1$ introduces Newtonian's intertrajectory coupling force
$F\backsim M^{2}L^{-2}$, where mass $M$ emerges as the dimensional coefficient
of proportionality.

It will be demonstrated how the mechanical principle of minimum action applied
to musculoskeletal system of animals involved in efficient locomotion may
provide basic patterns of biomechanical similarity.

\section{Minimum action in biomechanics}

During locomotion, chemical energy released by muscles and mechanical elastic
energy stored in body system is transformed into external and internal work
and partially lost as a heat. In the case of the off resonance human walking
[17], the small velocity-dependent frictional effects were accounted for in
the second order of perturbation theory, thereby generalizing the Lagrangian
formalism over weakly open systems.

During the muscle forced resonance walking and running (or flying) minimizing
energy consumption, the small damping effects restrict only the amplitude of
motion, \textit{i.e.}, \emph{stride length} $\Delta L$ (or stroke amplitude)
and \emph{muscle length change} $\Delta L_{m}$, but not the propagation
\emph{speed} $V=\Delta L/T$ and \emph{period} $T$, constrained geometrically
[14]. Likewise [17], frictional effects can be therefore neglected in the
equations of motion [14], on the first approximation. With the same precision,
the principle of mechanical similarity (\ref{1A}) provides%
\begin{align}
T^{-1}  &  =1/T_{ms}\varpropto\sqrt{E_{ms}}L_{m}^{-1}\text{, }V\backsim
V_{ms}\varpropto\sqrt{E_{ms}}\text{,}\nonumber\\
F  &  \backsim\Delta F\backsim F_{ms}\backsim\Delta F_{ms}=\varepsilon
_{m}A_{m}E_{ms}\text{, with }E_{ms}\varpropto(L_{m})^{s}\text{ and }%
g_{ms}\varpropto(L_{m})^{s-1}\text{,} \label{res}%
\end{align}
when presented in the linear-displacement body \ ($\Delta L\backsim L$) and
muscle ($\Delta L_{m}\backsim L_{m}$) approximation. Introducing in eq.
(\ref{res}) the force change $\Delta F$ for the body \emph{force output} $F$,
driving a given animal (of \emph{characteristic length} $L$,
\emph{cross-sectional area} $A$, and \emph{body mass} $M$) through the
environment, and the effective body rigidness, or longitudinal
\emph{stiffness} $K=\Delta F/\Delta L$, one also determines the natural
(resonant) \emph{cyclic frequency} $T^{-1}\backsim\sqrt{K/M}$ [1,7,8,17,18].
Since the animal locomotion is substantially muscular [1,3,18], the
\emph{muscle stiffness} $K_{m}=E_{m}A_{m}/L_{m}$ (of a muscle of length
$L_{m}$ and cross-sectional area $A_{m}$), controlled by the
geometry-independent muscle rigidity or \emph{elastic modulus} $E_{m}$ (ratio
of \emph{stress} $\sigma_{m}$ to \emph{strain }$\varepsilon_{m}$,
\textit{i.e.}, $(\Delta F_{m}/A_{m})/(\Delta L_{m}/L_{m})$) [7, 18], is also
under our consideration. To improve the integrative approach to animal
locomotion [1-18] via mechanical [19] and elastic strain [19, 20]
similarities, let us determine a muscle-force \emph{field} $g_{m}\equiv
F_{m}/m$, where the \emph{muscle mass} $m$ (or \emph{motor mass }[6]) is a
source of the \emph{active force output} $F_{m}$. Furthermore, the scaling
relations for physical quantities (shown in eq. (\ref{res}) by symbol
$\varpropto$) result from provided relations and constraints imposed by the
invariable \emph{body} \emph{density }$\rho$ ($=M/AL$) and \emph{muscle
density} $\rho_{m}$ ($m/A_{m}L_{m}$), all common in scaling biomechanics [1,7,9,18].

In this study, the intrinsic muscle modulus $E_{ms}$, substituting $E_{m}$ in
eq. (\ref{res}), describes a new dynamic degree of freedom characterizing
muscle ability of tuning to the resonance [15] in different locomotory gaits
distinguished by the single \emph{dynamic-state exponent} $s$.

\section{Results and discussion}

The \emph{steady-speed locomotion} for flight mode was first recognized by
Hill: "the frequencies of hovering birds are in inverse proportionality to the
cube roots of the weights, \textit{i.e.}, to the linear size" [2]. This
dynamic regime is pronounced in eq. (\ref{res}), taken with $s=0$, by the
propagation frequency $T^{-1}\backsim\sqrt{E_{m0}/\rho}L^{-1}$, contrasting
with the rigid-pendulum estimate $T_{pend}^{-1}\backsim\sqrt{g}L^{-1/2}$ ($g$
is gravitation field) [7,14]. Broadly speaking, Hill's observation plays the
role similar to Kepler's observation of third law for planets $T^{2}\varpropto
L^{3}$, following from eq. (\ref{1A}) with $s=-1$.

Hence, when the animal's body travels or cruises slowly for long distances [4]
with the constant optimum speed $V_{body}^{(\max)}$ $\backsim\sqrt
{E_{m0}^{(\max)}/\rho}$, invariant with body weight and frequency,\ or moves
throughout the terrestrial, air, or water environment resisting drag forces,
the legs, wings, and tails suggest to maintain constant elastic modulus
$E_{m0}^{(\max)}$ in \emph{slow muscles} responsible for the steady locomotion
[21]. Consequently, a constant functional intrinsic muscle stress
$\varepsilon_{m}E_{m0}$ is also predicted in eq. (\ref{res}) with $s=0$,
providing in turn constant safety factor (ratio of muscle strength to peak
functional stress), also expected by Hill [2]. These and other relevant
constraints of steady-speed locomotion are displayed in table 1.

.%

\begin{tabular}
[c]{||c|c|c|c|c|c||}\hline\hline
$s=0$ & Frequency & Length & \ Speed \  & \ \ \ Force\ \ \ \ \ \ \  &
Mass\\\hline\hline
\multicolumn{1}{||c|}{$\ \ T^{-1}$} & \multicolumn{1}{|r|}{$T^{-1}$} &
\multicolumn{1}{|r|}{$\rho^{{\small -}\frac{1}{2}}E_{{\small 0}}^{\frac{1}{2}%
}\cdot L^{{\small -1}}$} & \multicolumn{1}{|r|}{$\rho^{{\small -}\frac{1}{4}%
}E_{{\small 0}}^{\frac{1}{4}}\cdot V^{-\frac{1}{2}}$} &
\multicolumn{1}{|r|}{$F^{0}$} & \multicolumn{1}{|r||}{$\rho^{{\small -}%
\frac{1}{6}}E_{{\small 0}}^{{\small -}\frac{1}{2}}\cdot M^{{\small -}\frac
{1}{3}}$}\\\hline
\multicolumn{1}{||c|}{$\Delta L,L$} & \multicolumn{1}{|r|}{$\rho
^{{\small -}\frac{1}{2}}E_{{\small 0}}^{\frac{1}{2}}\cdot T$} &
\multicolumn{1}{|r|}{$L$} & \multicolumn{1}{|r|}{$\rho^{{\small -}\frac{1}{4}%
}E_{{\small 0}}^{\frac{1}{4}}\cdot V^{\frac{1}{2}}$} &
\multicolumn{1}{|r|}{$F^{0}$} & \multicolumn{1}{|r||}{$\rho^{{\small -}%
\frac{1}{3}}\cdot M^{\frac{1}{3}}$}\\\hline
\multicolumn{1}{||c|}{$V^{(\max)}$} & \multicolumn{1}{|r|}{$\rho
^{{\small -}\frac{1}{2}}E_{{\small 0}}^{\frac{1}{2}}\cdot T^{0}$} &
\multicolumn{1}{|r|}{$\rho^{{\small -}\frac{1}{2}}E_{{\small 0}}^{\frac{1}{2}%
}\cdot L^{0}$} & \multicolumn{1}{|r|}{$\rho^{{\small -}\frac{1}{2}%
}E_{{\small 0}}^{\frac{1}{2}}$} & \multicolumn{1}{|r|}{$\rho^{{\small -}%
\frac{1}{2}}E_{{\small 0}}^{\frac{1}{2}}\cdot F^{0}$} &
\multicolumn{1}{|r||}{$\rho^{{\small -}\frac{1}{2}}E_{{\small 0}}^{\frac{1}%
{2}}\cdot M^{{\small 0}}$}\\\hline
\multicolumn{1}{||c|}{$K_{body}^{(\max)}$} & \multicolumn{1}{|r|}{$\rho
^{\frac{1}{2}}E_{{\small 0}}^{\frac{1}{2}}A\cdot T^{-1}$} &
\multicolumn{1}{|r|}{$E_{{\small 0}}A\cdot L^{-1}$} &
\multicolumn{1}{|r|}{$\rho^{\frac{1}{4}}E_{{\small 0}}^{-\frac{1}{4}}\cdot
V^{-\frac{1}{2}}$} & \multicolumn{1}{|r|}{$L^{-1}\cdot F$} &
\multicolumn{1}{|r||}{$\rho^{{\small -}\frac{1}{3}}E_{0}\cdot M^{\frac{1}{3}}%
$}\\\hline
\multicolumn{1}{||c|}{$\sigma_{slow}^{(\max)}$} &
\multicolumn{1}{|r|}{$\varepsilon_{m}E_{0}\cdot T^{0}$} &
\multicolumn{1}{|r|}{$\varepsilon_{m}E_{0}\cdot L^{0}$} &
\multicolumn{1}{|r|}{$\varepsilon_{m}E_{0}\cdot V^{0}$} &
\multicolumn{1}{|r|}{$\varepsilon_{m}E_{0}\cdot F^{0}$} &
\multicolumn{1}{|r||}{$\varepsilon_{m}E_{0}\cdot m^{{\small 0}}$}\\\hline
\multicolumn{1}{||c|}{$F_{slow}^{(\max)}$} & \multicolumn{1}{|r|}{$\varepsilon
_{m}E_{0}A_{m}\cdot T^{0}$} & \multicolumn{1}{|r|}{$\varepsilon_{m}E_{0}%
A_{m}\cdot L^{0}$} & \multicolumn{1}{|r|}{$\varepsilon_{m}E_{0}A_{m}\cdot
V^{0}$} & \multicolumn{1}{|r|}{$\varepsilon_{m}^{(\max)}E_{0}A_{m}$} &
\multicolumn{1}{|r||}{$\rho_{{\small m}}^{{\small -}\frac{2}{3}}%
\varepsilon_{m}E_{0}\cdot m^{\frac{2}{3}}$}\\\hline
\end{tabular}

Table 1. Mechanical characteristics of body system and slow individual muscles
in\textrm{ }the steady-motion dynamic states $s=0$ prescribed by the principle
of minimum muscular action in eq. (\ref{res}). \emph{Abbreviation}:
$E_{0}=E_{m0}^{(\max)}$.

.

The constant maximum propulsive force $F_{body}^{(\max)}\backsim E_{m0}%
^{(\max)}A$, equilibrating all drag forces via slow muscles, \textit{i.e.},
$F_{drag}^{(\max)}\backsim F_{slow}^{(\max)}$ shown in table 1, was first
documented by Alexander as the peak body force $F_{body}^{(\exp)}\varpropto
M^{2/3}$ [10] exerted on the environment by running, flying, and swimming
animals ranged over nine orders of body mass. More recently, the slow-fiber
force output $F_{slow}^{(\max)}\varpropto m^{2/3}$ (table 1) was revealed [6]
by statistical regression method in both biological and human-made \emph{slow
motors}. The underlying muscle longitudinal field "caused by intrinsic muscle
quantity (here associated with $E_{m0}$), equally stimulated electrically and
by the nervous system" [2] decreases linearly with the distance $r$:
$g_{slow}^{(\max)}(r)\thickapprox E_{m0}^{(\max)}\varepsilon_{m}^{(\max)}%
/\rho_{m}r$, where $\varepsilon_{m}^{(\max)}$ is nearly isometric strain, as
follows from eq. (\ref{res}) with $s=0$.

The resonant efficient locomotion broadly prescribes a concerted behavior
synchronized in time and coordinated in displacements and forces of the body's
appendages. Consequently, the muscle \emph{duty factor} $\beta_{m}=\Delta
t_{m}/T$, where $\Delta t_{m}$ is timing of the muscle lengthening/shortening
$\Delta L_{m}$, is constant, besides the body-mass invariable \emph{Strouhal
number} $St=\Delta L/VT$, explaining the tail and wing oscillations in
swimmers and flyers [22]. At maximum propulsive efficiency of cruising
dolphins, birds, and bats, it was observed as $St_{cruis}\thickapprox0.3$ [4].

The steady-speed locomotion state also was remarkably established in hovering
flying motors via the wing frequencies $1/T^{(\exp)}\varpropto M^{-1/3}$ [23],
as predicted in table 1. However, departures from Hill's findings rationalized
here by the dynamic-state exponent $s=0$ were also debated [24]. For example,
it was claimed [7] that Hill's maximal optimum speeds are in sharp
disagreement with the peak trot-gallop \emph{crossover speeds} $V_{cross}%
^{(\exp)}$ measured in quadrupeds [12]. The same could refer to the bipeds
[11]. However, as can be seen from the proper empirical data $1/T^{(\exp
)}\varpropto M^{-0.178}$ [11] and $1/T^{(\exp)}\varpropto V_{cross}^{-1}$
$\varpropto M^{-0.145}$ [7,12], the measured stride frequencies indicate
observations of another kind of mechanical similarity attributed to the
\emph{non-steady dynamic state} $s=1$, prescribed in eq. (\ref{res}) through
the mass-dependent muscle modulus $E_{m1}\varpropto L_{m}\varpropto M^{1/3}$.

The minimum muscle action of legs in fast running rats, wallaby, dog, goat,
horse, and human was indirectly revealed through the mechanical similarity
derived with the help of \emph{leg spring model} [8], providing the stride
frequency $T^{-1}\backsim\Delta t_{leg}^{-1}\varpropto M^{-0.19}$, stride
length $\Delta L\varpropto M^{0.30}$, model-body length $L_{leg}\varpropto
M^{0.34}$, body stiffness $K_{leg}^{(\max)}\varpropto M^{0.67}$, and body
force output $F_{leg}^{(\max)}\varpropto M^{0.97}$. Relations between the
quantities underlying these findings are discussed below and summarized in
table 2.

.%

\begin{tabular}
[c]{||c|c|c|c|c|c||}\hline
$s=1$ & Frequency & Length & \ Speed \  & \ \ \ Force\ \ \ \ \ \ \  &
Mass\\\hline\hline
\multicolumn{1}{||c|}{$T^{-1}$} & \multicolumn{1}{|r|}{$T^{-1}$} &
\multicolumn{1}{|r|}{$g_{1}^{\frac{1}{2}}\cdot L^{-\frac{1}{2}}$} &
\multicolumn{1}{|r|}{$g_{1}\cdot V^{-1}$} & \multicolumn{1}{|r|}{$(\rho
g_{1}^{2}A)^{{\small -}\frac{1}{2}}\cdot F^{\frac{1}{2}}$} &
\multicolumn{1}{|r||}{$\rho^{\frac{1}{6}}g_{1}^{\frac{1}{2}}\cdot
M^{{\small -}\frac{1}{6}}$}\\\hline
\multicolumn{1}{||c|}{$\Delta L,L$} & \multicolumn{1}{|r|}{$g_{1}\cdot T^{2}$}
& \multicolumn{1}{|r|}{$L$} & \multicolumn{1}{|r|}{$g_{1}^{-1}\cdot V^{2}$} &
\multicolumn{1}{|r|}{$(\rho g_{1}A)^{{\small -1}}\cdot F$} &
\multicolumn{1}{|r||}{$\ \rho^{{\small -}\frac{1}{3}}\cdot M^{\frac{1}{3}}$%
}\\\hline
\multicolumn{1}{||c|}{$V_{cross}^{(\max)}$} & \multicolumn{1}{|r|}{$g_{1}\cdot
T$} & \multicolumn{1}{|r|}{$g_{1}^{\frac{1}{2}}\cdot L^{\frac{1}{2}}$} &
\multicolumn{1}{|r|}{$V$} & \multicolumn{1}{|r|}{$(\rho A)^{{\small -}\frac
{1}{2}}\cdot F^{\frac{1}{2}}$} & \multicolumn{1}{|r||}{$\rho^{{\small -}%
\frac{1}{6}}g_{1}^{\frac{1}{2}}\cdot M^{\frac{1}{6}}$}\\\hline
\multicolumn{1}{||c|}{$K_{body}^{(\max)}$} & \multicolumn{1}{|r|}{$\rho
g_{1}A\cdot T^{0}$} & \multicolumn{1}{|r|}{$\rho g_{1}A\cdot L^{0}$} &
\multicolumn{1}{|r|}{$\rho g_{1}A\cdot V^{0}$} & \multicolumn{1}{|r|}{$\rho
g_{1}A\cdot F^{0}$} & \multicolumn{1}{|r||}{$\rho^{{\small -}\frac{1}{3}}%
g_{1}\cdot M^{\frac{2}{3}}$}\\\hline
\multicolumn{1}{||c|}{$\sigma_{fast}^{(\max)}$} & \multicolumn{1}{|r|}{$\rho
_{m}g_{1}^{2}\cdot T^{2}$} & \multicolumn{1}{|r|}{$\rho_{m}g_{1}\cdot L_{m}$}
& \multicolumn{1}{|r|}{$\rho_{m}\cdot V^{2}$} & \multicolumn{1}{|r|}{$A_{m}%
^{-1}\cdot F_{m}$} & \multicolumn{1}{|r||}{$\rho_{m}^{\frac{2}{3}}g_{1}\cdot
m^{\frac{1}{3}}$}\\\hline
\multicolumn{1}{||c|}{$F_{fast}^{(\max)}$} & \multicolumn{1}{|r|}{$\rho
_{m}g_{1}^{2}A_{m}\cdot T^{2}$} & \multicolumn{1}{|r|}{$\rho_{m}g_{1}%
A_{m}\cdot L$} & \multicolumn{1}{|r|}{$\rho_{m}A_{m}\cdot V^{2}$} &
\multicolumn{1}{|r|}{$F_{m}$} & \multicolumn{1}{|r||}{$g_{1}\cdot m$%
}\\\hline\hline
\end{tabular}

Table 2. Mechanical characterization of body of animals and fast muscles in
physiologically equivalent non-steady states $s=1$ prescribed by eq.
(\ref{res}). \emph{Abbreviation}: \medskip$g_{1}=g_{m1}$.

.

In accord with table 2, the equilibration of the air drag by wings of flapping
birds is manifested by the observed wing frequencies $1/T^{(\exp)}\varpropto
M^{-1/6}$ [23]. Moreover, the mechanical similarity between animals resisting
air, ground, and water friction forces was demonstrated via the energy cost
minimization [9], where the spatiotemporal correlations $V\varpropto L^{1/2}$
($\varpropto M^{1/6}$) were critically explored on \textit{ad hoc} basis.

When the non-steady locomotion conditions associated with the physiologically
equivalent (or transient-equilibrium [19]) states $s=1$ are applied to
individual fast-twitch-fiber muscles controlling fast gaits [21], the muscle
field is apparently uniform and likely universal [6]. Indeed, the body force
field $F_{body}^{(\max)}/M\thickapprox3g$ was first observed via the maximum
force output in fast trotting and hopping quadrupeds [8]. Later, mass-specific
force output $g_{m1}^{(\exp)}$ was empirically established [6] for locomotory
individual muscles associated with \emph{fast motors} in running, flying, and
swimming animals. One therefore infers that the gravitation field $g$ is not
crucial in fast running modes, as proposed in [9]. Moreover, the principle of
minimum muscular action suggests that fast muscles may generate force into the
whole muscle bulk [25] maintaining constant body stiffness (table 2), unlike
the constant pressure characteristic of steady gaits (table 1). In other
words, the fast muscles are not simple passive springs [3,26], attributed to
$s=2$ and having length-independent period, but are complex systems being able
to activate fibres in both parallel and series. Maintaining the uniform muscle
force field $g_{m1}$, the \emph{Froude number }($Fr=V/\sqrt{gL}$ [1]) must be
mass-invariable, for both muscle system ($Fr_{fast}\backsim\sqrt{g_{m1}/g}$)
and body system, apart from the corresponding Strouhal number. For fast
running gaits in mammals, $Fr_{run}^{(\exp)}\thickapprox1.5$ and
$St_{run}^{(\exp)}\thickapprox0.4$ [8].

\section{Conclusion}

The main goal of this letter is to demonstrate how the complex biological
phenomenon of mechanical similarity in animal locomotion allows to be
rationalized and formulated as a predictive, quantitative framework. It has
been shown how the fundamental physical principle of minimum action applied to
locomotory muscles via intrinsic elastic moduli quantifies amazing
similarities established empirically between maximal speeds, frequencies,
forces, and other relevant mechanical characteristics of animals locomoting in
a certain gait. Naturally operating the softness of legs, wings, and tails,
the efficient runners, flyers, and swimmers are shown to maintain constant
Strouhal number via the universal constant muscle pressure, when traveling or
cruising at steady speeds. When acting quickly at higher speeds, escaping from
predators, or when hunting, the successful runners, flyers, and swimmers
appear to maintain the universal field in the whole bulk of fast muscles, at
least at crossover speeds. This uniform field eventually results in the
bodyweight depending, fixed muscle stiffness and universal Froude and Strouhal
numbers. The provided from first principles study illuminates and supplements
a wide spectrum of reliable empirical findings in walking and running bipeds
[3,11], trotting and galloping quadrupeds [6-9,12]; hovering and flapping
birds [2-4,10,11], bats, and insects [3,4,9]; undulating and tail-beating fish
[2-4,9,10], dolphins [2,4], sharks [4], and whales [2].

On the other hand, the study of muscle characteristics, including obtained
scaling relations to muscle and body mass, is limited by the
linear-displacement muscle approximation. It can been shown however that the
top speeds attributed to limiting animal performance [19,24] cannot be
achieved by the linear-strain elastic muscle fields. The consequences of
application of the minimum action to specific fast locomotory muscles
structurally adapted to a certain mechanical activity, such as motor, brake,
or strut functions [3] prescribed by non-linear elastic effects [25] will be
discussed elsewhere.

.

\textbf{Acknowledgments}. Financial support by the national agency CNPq is acknowledged.

.

\textbf{References}

1. Alexander R. McN. \textit{Principles of animal locomotion} (Princeton
University Press, Princeton and Oxford) 2002 pp.53-67

2. Hill A. V. 1950 The dimensions of animals and their muscular dynamics
\textit{Sci. Progr.} \textbf{38} 209-230

3. Dickinson M. H., Farley C. T., Full J. R., Koehl M. A. R., Kram R. and
Lehman S. 2000 How animals move: an integrative view \textit{Science}
\textbf{288} 100-106

4. Taylor G. K., Nudds R. L. and Thomas A. L. 2003 Flying and swimming animals
cruise at a Strouhal number tuned for high power efficiency \textit{Nature}
\textbf{425} 707-710

5. Darveau C. A., Suarez R. K., Andrews R. D. and Hochachka P. W. 2002
Allometric cascade as a unifying principle of body mass effects on metabolism
\textit{Nature} \textbf{417} 166--170

6. Marden J. H. and Allen L. R. 2002 Molecules, muscles, and machines:
universal performance characteristics of motors \textit{Proc. Natl. Acad. Sci.
USA} \textbf{99} 4161-4166

7. McMahon T. A. 1975 Using body size to understand the structural design of
animals: quadrupedal locomotion \textit{J. Appl. Physiol.} \textbf{39} 619-627

8. Farley C. T., Glasheen J. and McMahon T. A. 1993 Running springs: speed and
animal size \textit{J. Exp. Biol.} \textbf{185} 71-86

9. Bejan A. and Marden J. H. 2006 Unifying constructal theory for scale
effects in running, swimming and flying \textit{J. Exp. Biol}. \textbf{209} 238-248

10. Alexander R. McN. 1985 The maximum forces exerted by animals \textit{J.
Exp. Biol.} \textbf{115} 231-238

11. Gatezy S. M. and Biewener A.A. 1991 Bipedal locomotion: effects of speed,
size and limb posture in birds and animals \textit{J. Zool. Lond}.
\textbf{224} 127-147

12. Heglund N., McMahon T. A. and Taylor C. R. 1974 Scaling stride frequency
and gait to animal size: mice to horses \textit{Science} \textbf{186} 1112-1113

13. Cressey D. 2008 Moving forward together \textit{Nature} doi:10.1038/news.2008.1268

14. Ahlborn B. K. and Blake R.W. 2002 Walking and running at resonance
\textit{Zoology} \textbf{105} 165--174

15. Ahlborn B. K., Blake R.W. and Megill W. M. 2006 Frequency tuning in animal
locomotion \textit{Zoology} \textbf{109} 43--53

16. Landau L. D. and Lifshitz E. M. \textit{Mechanics }(Pergamon Press, 3rd
ed., Oxford) 1976 section 10

17. Kokshenev V. B. 2004 Dynamics of human walking at steady speeds
\textit{Phys. Rev. Lett.} \textbf{93} 208101--208105

18. McMahon T. A. \textit{Muscles, reflexes, and locomotion} (Princeton
University Press, Princeton and New Jersey) 1984

19. Kokshenev V. B. 2007 New insights into long-bone biomechanics: Are limb
safety factors invariable across mammalian species? \textit{J. Biomech.}
\textbf{40} 2911-2918

20. Rubin C. T. and Lanyon L. E. 1984 Dynamic strain similarity in
vertebrates; an alternative to allometric limb bone scaling \textit{J. Theor.
Biol}. \textbf{107} 321--327

21. Rome L .C., Funke R. P., Alexander R. McN., Lutz G., Aldridge H., Scott F.
and Freadman M. 1988 Why animals have different muscle fibre types
\textit{Nature} \textbf{335} 824-829

22. Whitfield J.2003 One number explains animal flight\textit{Nature} doi:10.1038/news031013-9

23. Ellington C. P. 1991 Limitations on animal flight performance \textit{J.
Exp. Biol.} \textbf{160} 71-91

24. Jones J. H. and Lindstedt S. 1993 Limits of maximal performance
\textit{Annu. Rev. Physiol.} \textbf{55} 547-569

25. Kokshenev V. B. 2008 A force-similarity model of the activated muscle is
able to predict primary locomotor functions \textit{J. Biomech. }\textbf{41} 912--915

26. Lindstedt S. L., Reich T. E., Keim P. and LaStayo P. C. 2002 Do muscle
functions as adaptable locomotor springs? \textit{J. Exp. Biol.} \textbf{205} 2211-2116

\end{document}